\documentclass[useAMS,usenatbib]{mn2e}
\usepackage{graphicx}
\usepackage{epsfig}
\DeclareGraphicsExtensions{.pdf,.png,.jpg}
\usepackage{amssymb}
\usepackage{amsmath}
\usepackage{rotating}
\usepackage{lscape}
\usepackage{array}
\def\hi{H\,{\sc i} }

\def\kms{km~s$^{-1}$}
\def\msun{M$_{\odot}$}

\def\ks{K$_{\rm S}$}

\long\def\symbolfootnote[#1]#2{\begingroup%
\def\thefootnote{\fnsymbol{footnote}}\footnote[#1]{#2}\endgroup} 
\usepackage{epsfig}
\usepackage{color}\DeclareGraphicsExtensions{.pdf,.png,.jpg}
\usepackage[]{subfigure}
   
\def\beq{\begin{equation}} 
\def\eneq{\end{equation}} \def\bea{\begin{eqnarray}} \def\enea{\end{eqnarray}}

\providecommand{\degree}{\ensuremath{^\circ}}

\title[The Distribution of Mass in the Orion Dwarf Galaxy]{The Distribution of Mass in the Orion Dwarf Galaxy}
\author[Frusciante et al.]{N. Frusciante$^{1}$\thanks{E-mail:
    noemi.frusciante@sissa.it}, P. Salucci$^{1}$\thanks{E-mail:
    paolo.salucci@sissa.it}, D. Vernieri$^{1}$\thanks{E-mail:
    daniele.vernieri@sissa.it}, J. M. Cannon$^{2}$\thanks{E-mail:
    jcannon@macalester.edu}, E. C. Elson$^{3}$\thanks{E-mail:
    ed.elson@uwa.edu.au} \\
$^{1}$SISSA/ISAS, International School for Advanced Studies, Via
  Bonomea 265, 34136, Trieste, Italy; \\
INFN, Sezione di Trieste, Via Valerio 2, 34127, Trieste, Italy \\
$^{2}$Department of Physics \& Astronomy, Macalester College, 1600
Grand Avenue, Saint Paul, MN 55105, USA\\
$^{3}$International Centre for Radio Astronomy Research, The
University of Western Australia, M468, 35 Stirling Highway,
\\ Crawley, WA 6009, Australia}

\begin{document}

\date{\today}

\pagerange{\pageref{firstpage}--\pageref{lastpage}} \pubyear{2011}

\maketitle

\label{firstpage}

\begin{abstract}
Dwarf galaxies are good candidates to investigate the nature of Dark
Matter, because their kinematics are dominated by this component
down to small galactocentric radii. We present here the results of
detailed kinematic analysis and mass modelling of the Orion dwarf
galaxy, for which we derive a high quality and high resolution
rotation curve that contains negligible non-circular motions and we
correct it for the asymmetric drift.  Moreover, we leverage the
proximity (D $=$ 5.4 kpc) and convenient inclination (47\degree) to
produce reliable mass models of this system.  We find that the
Universal Rotation Curve mass model (Freeman disk $+$ Burkert
halo $+$ gas disk) fits the observational data accurately.  In contrast, the
NFW halo + Freeman disk $+$ gas disk mass model is unable to reproduce the observed
Rotation Curve, a common outcome in dwarf galaxies.  Finally, we
attempt to fit the data with a MOdified Newtonian Dynamics (MOND)
prescription. With the present data and with the present assumptions on distance, stellar mass, constant inclination and reliability of the gaseous mass,  the MOND ``amplification''  of  the baryonic component appears to be too small to mimic the required ``dark
component''.  The Orion dwarf reveals a cored DM density distribution and a
possible tension between observations and the canonical MOND formalism.
\end{abstract}

\begin{keywords}
Dark Matter; Galaxy: Orion dwarf; Mass profiles 
\end{keywords}

\section{Introduction}

The measurement of the Rotation Curves (RCs) of disk galaxies is a
powerful tool to investigate the nature of Dark Matter (DM), including
its content relative to the baryonic components and their
distributions. In particular, dwarf galaxies are good candidates to
reach this aim as their kinematics are generally dominated by the dark
component, down to small galactocentric radii \citep{persalstel,
  gentile04, blok, oh, salucci08}. This leads to a reliable measurement of the
dynamical contribution of the DM to the RC and hence of its density
profile. Therefore, a dwarf galaxy like the Orion dwarf provides us
with an important test as to whether DM density profiles arising in
 $\Lambda$ Cold Dark Matter ($\Lambda$CDM) numerical simulations \citep{nfw} are compatible with
those detected in actual DM halos around galaxies. Let us comment that NFW profile arises from pure  N-body DM
simulations. It is well known that, as effect of the baryonic infall in the cosmological DM halos and of  the subsequest  process of stellar disk formation,  shallower profiles of the DM halo may arise (see \citet{governato, governato12, maccio}).

\begin{figure*}
\begin{centering}
\includegraphics[angle=0,width=18truecm]{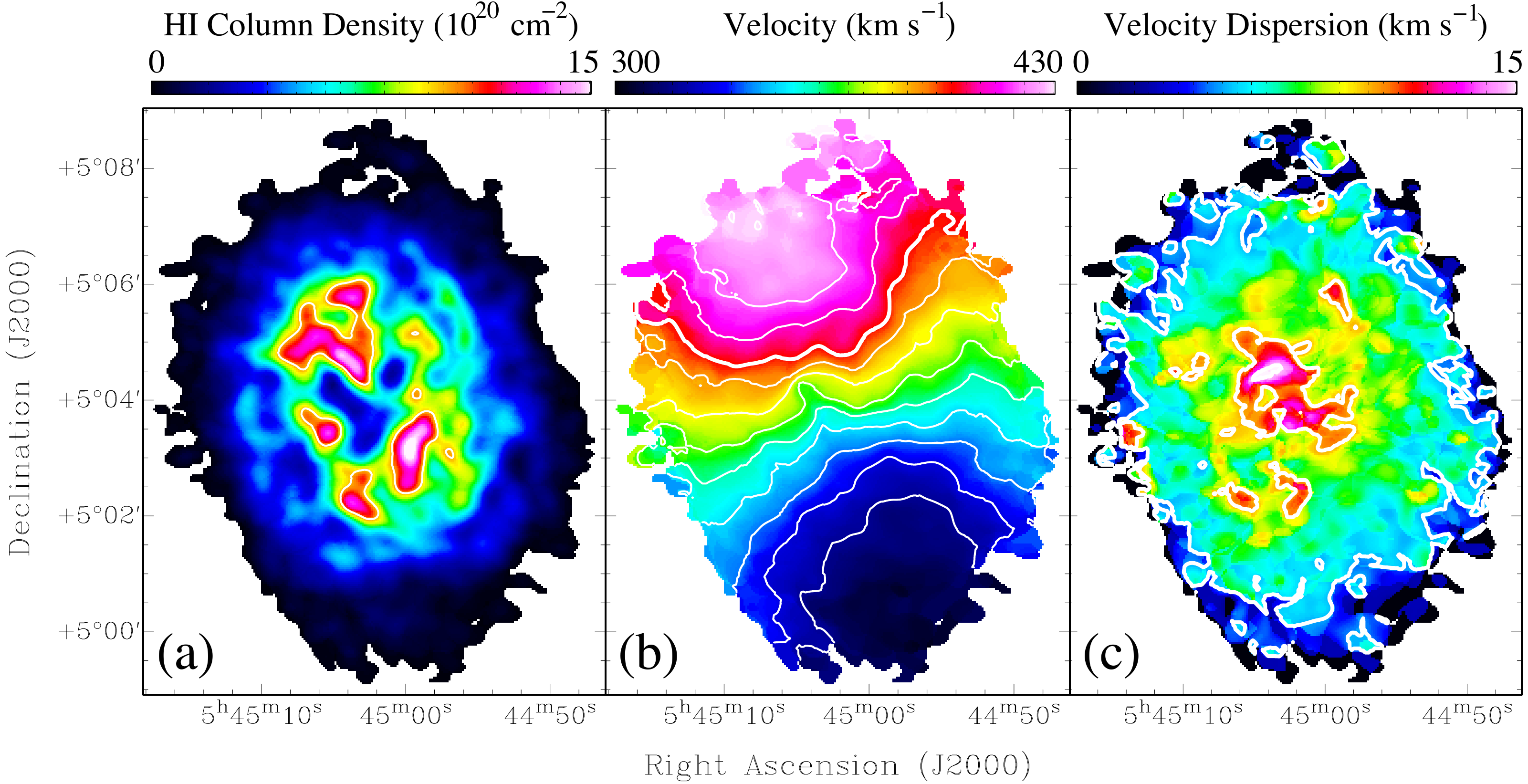}
\caption{Comparison of the \hi\ column density distribution ({\it a}),
  the intensity-weighted velocity field ({\it b}), and the intensity-weighted
  velocity dispersion ({\it c}); the beam size is 20\arcsec.  The
  contour in ({\it a}) is at the 10$^{21}$ cm$^{-2}$ level; the contours
  in ({\it b}) show velocities between 320 and 420 \kms, separated by
  10 \kms, and the thick white line corresponds to the 400 \kms contour; the contours in ({\it c}) are at the 5, 10, 15
  \kms\ levels.}
\label{figcap3}	
\end{centering}
\end{figure*}  

Recent studies of the RCs of dwarf galaxies have tested the NFW
scenario. It is now clear that kinematic data are better fitted by a
DM halo with a constant density core (e.g. \citet{borriello,bosch,blok,weldrake,gentile04,simon,oh}), than
by one that is centrally peaked.  One specific example is DDO 47,
whose velocity field is clearly best fitted if the DM halo is cored;
moreover, its (small) detected non-circular motions cannot account for
the discrepancy between data and the NFW predictions
\citep{gentile05}. \\
The present investigation examines the DM content of the Orion dwarf
galaxy. This nearby system harbors an extended \hi\ disk, and thus
provides us with an important test of the above paradigm. As we show
below, the Orion dwarf is one of the few known galaxies whose
kinematics {\it unambiguously} point towards a cored profile.  This
system is thus critically important for investigating the nature of
the DM particle and of the evolution of DM halos. \\
MOND accounts for the evidence that RCs of spiral galaxies are
inconsistent with the corresponding distribution of the luminous
matter \citep{milgroma,milgromb}.  Rather than postulating the
existence of a dark halo made by massive collisionless elementary
particles, this scenario advocates that the gravitational force at low
accelerations leaves the standard Newtonian regime to enter a very different one. Historically MOND has generally been successful in
reproducing the RCs of spiral galaxies with only the (observed)
luminous matter (e.g. \citet{sanders96,sanders98,sanders07}).
However, cases of tension between data and the MOND formalism do exist \citep{gentile04}. \\
It is important to stress that in order to derive the DM density
profile or to test the MOND formalism, we must know the distribution
of the ordinary baryonic components, as well as have reliable
measurements of the gas kinematics.  For the Orion dwarf, 21-cm
\hi\ surface brightness and kinematics have recently been published
\citep{Cannon}: their analysis provides a high quality, high
resolution RC, that, in addition, can be easily corrected for
asymmetric drift and tested for non-circular motions.  This galaxy is
a very useful laboratory in that a simple inspection of the RC ensures
us that it shows a large mass discrepancy at all radii.  Moreover, the
baryonic components are efficiently modeled (i.e., no stellar bulge is
evident and the stellar disk shows a well-behaved exponential profile,
see \citet{vaduvescu}). The distance to the galaxy, which is critical for an unambiguous test of MOND \citep{sanders02}, is estimated to be 5.4$\pm$1.0 Mpc \citep{vaduvescu}. It is important to stress that the distance of the Orion dwarf remains a significant source of uncertainty.  \citet{vaduvescu} estimate the distance using the brightest stars method.  The intrinsic uncertainty in this technique may allow a distance ambiguity much larger than the formal errors estimated by \citet{vaduvescu}, because in their work this method yields a scatter as large as $50\%$ in distance. Finally, the system's inclination (47\degree) is
kinematically measured (see Section (\ref{3.1})) and is high enough to not affect the
estimate of the circular velocity.  The properties described above
make the Orion dwarf galaxy an attractive candidate to determine the
underlying gravitational potential of the galaxy.

This paper is organized as follows.  In Sec. 2 we present the stellar
surface photometry.  In Sec. 3, the \hi\ surface density and
kinematics data are presented and discussed; we also provide the analysis of
possible non-circular motions of the neutral gas.  In Sec. 4 we model
the RC in the stellar disk using a cored/cusped halo framework. In
Sec. 5 we test the Orion kinematics against the MOND formalism. Our
conclusions are given in Sec. 6.
  
\section{Stellar Photometry}

Following the discussion in \citet{Cannon}, the underlying stellar
mass in the Orion dwarf is estimated using the near-infrared (IR)
photometry (J and \ks\ bands) presented by \citet{vaduvescu}.  Those
authors find (J\,$-$\,\ks) $=$ $+$0.80 and a total \ks\ magnitude of
$+$10.90. When comparing to models (see below) we assume that the
color difference between K and \ks\ is negligible; further, we assume
L$_{\rm K,\,\odot} = +$3.33 \citep{cox,bessel}.  Accounting for
extinction, the total K-band luminosity of the Orion dwarf is
$\sim$3.5\,$\times$\,10$^8$ L$_{\odot}$. The mass of the stellar
component was estimated by \citet{Cannon} to be
(3.7\,$\pm$\,1.5)\,$\times$\,10$^{8}$ \msun. % The implied ratio of
% (M$_D$/L$_K$)$\simeq$ 1 is in very good agreement with the (M$_D$/L$_I$)$\simeq$ 0.5 that we obtain by extrapolating at low luminosities the
%relationship found by \citet{salucci08}. 
 The stellar surface brightness profile is well fitted by an exponential thin disk, with a
scale length of $R_D$= 25\arcsec\ $\pm$ 1\arcsec\ (equivalent to 1.33 $\pm$ 0.05
kpc at the adopted distance). Moreover, there are no departures from
an exponential profile that would be indicative of a prominent central
bulge.

\begin{figure*}
\begin{centering}
\includegraphics[angle=0,width=9truecm]{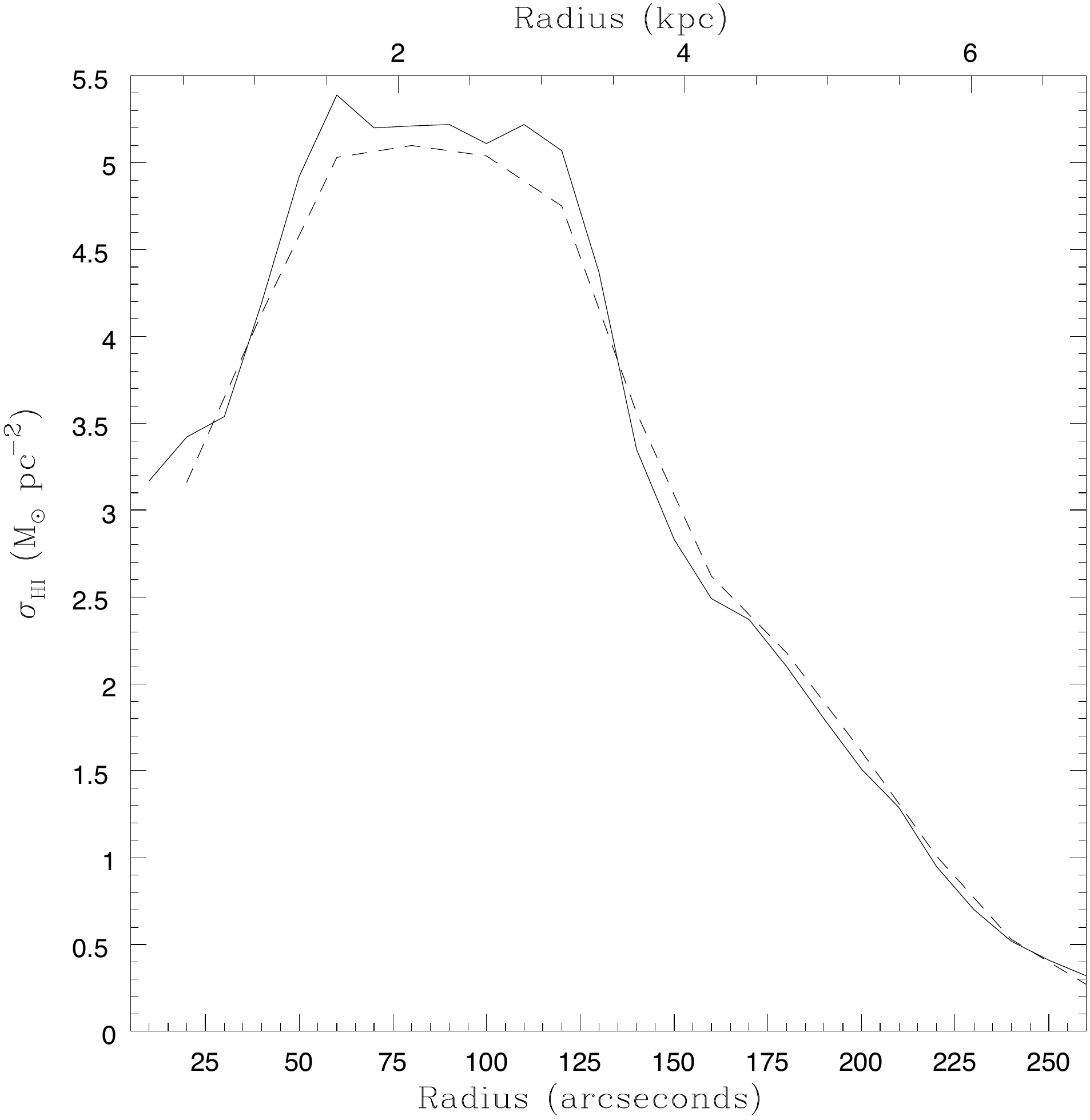}
\caption{Radially averaged \hi\ mass surface density profiles of the
  Orion dwarf, created by averaging \hi\ emission in concentric rings
  emanating from the dynamical center found in our RC
  analysis. The solid/dotted lines were created from the 10\arcsec\ /20\arcsec\ resolution
  images.}
\label{reswri}	
\end{centering}
\end{figure*}

\section{HI surface density and kinematics}

\hi\ spectral line imaging was acquired with the {\it Very Large
  Array} and presented in \citet{Cannon}. We refer the reader to that
work for a full discussion of the data handling, and we summarize
salient details here.  The final data cubes have a circular beam size
of 20\arcsec, with a 3\,$\sigma$ \hi\ column density sensitivity of
N$_{\rm HI}$= 1.5 $\times$10$^{19}$\,cm$^{-2}$. The first three moment
maps (i.e.  the integrated \hi\ intensity, the velocity field, and the
velocity dispersion) are shown in Figure (\ref{figcap3}). \\
The neutral gas disk of the Orion dwarf shows rich morphological and
kinematic structure at this physical resolution.  The outer disk
contains tenuous \hi\ gas, but column densities rise above the
5\,$\times$\,10$^{20}$ cm$^{-2}$ level at intermediate radii.  There
is plentiful high-column density ($>$10$^{21}$ cm$^{-2}$)
\hi\ throughout the disk. The more or less parallel iso-velocity contours at inner radii are indicative of linear rotation (although almost certainly not solid body) and the curving of the outer contours suggests that the outer rotation curve has a fairly constant velocity.
The outer disk contours show no evidence for a decrease in rotational velocity at large radii.  In the central
regions of the disk, however, some \hi\ ``holes'' or ``depressions''
manifest a pronounced kink in these contours (consider the contours at
370\,$\pm$\,20 \kms).  The intensity weighted velocity dispersion
averages to $\sim$7-8 \kms\ throughout the disk, although the
innermost regions show dispersions above 10 \kms. \\
The total \hi\ flux integral, proportional to the \hi\ disk mass, was
found to be 50.3\,$\pm$\,5.1 Jy\,km s$^{-1}$, a value somewhat lower than the
single-dish flux measure of 80.6$\pm$7.72 Jy\,km s$^{-1}$ by
\citet{springob05}; the difference may arise from the lack of short
interferometric spacings that provide sensitivity to diffuse
structure.  The total \hi mass is found to be M$_{\rm HI}$ $=$
(3.5$\pm$0.5)\,$\times$\,10$^{8}M_\odot$. After applying the usual
35\% correction for Helium and molecular material, we adopt M$_{\rm gas}$ $=$
(4.7$\pm$0.7)\,$\times$\,10$^{8}M_\odot$ as the total gas mass. \\
In Figure (\ref{reswri}) we plot the 10\arcsec\ /20\arcsec\ resolution
\hi\ surface density, throughout the gas disk. A simple fit (valid
out to the last measured point and for the scope of this work) yields:
\beq 
\mu_{\rm HI}(r)= \frac{-0.263 r^3+1.195 r^2+3.094 r+18.549}{0.154 r^3-1.437 r+6.703} \,\, M_\odot/pc^2, 
\eneq 
\noindent where $r$ is in kpc.  The related fitting uncertainty on $\mu_{\rm HI}(r)$ is
about $20 \%$.  %The \hi\ disk is significantly larger than the
%(exponential) stellar disk (see {Cannon et  al. 2010}\nocite{Cannon}) so that we can investigate the outer
%regions of Orion dwarf. 
Figure (\ref{reswri}) shows that the
\hi\ surface density rises from the center of the galaxy, reaches a
maximum, and then declines exponentially.  At the last measured point,
i.e.  out to $\sim$ 7 kpc, the profile  has  almost (though not completely) reached the edge of the \hi\ disk and rapidly converges to zero. Note that, in Newtonian gravity, the outer gaseous disk
contributes in a negligible way to the galaxy total gravitational
potential.

\subsection{The Circular Velocity}
\label{3.1}
\begin{figure*}
\begin{centering}
\includegraphics[angle=0,height=7truecm,width=12truecm]{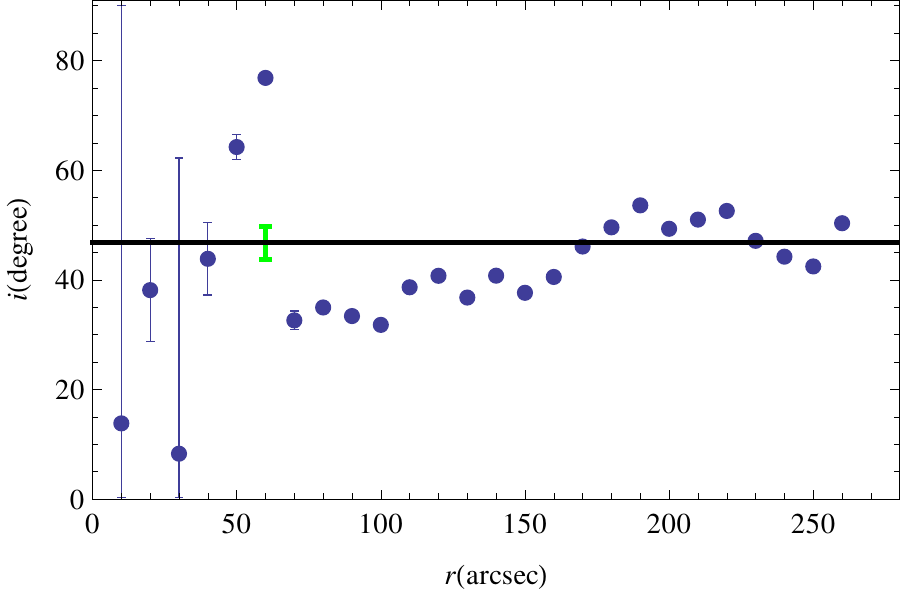}
\caption{Inclination best fit (thick Black line) for high resolution ($10''$) data (filled Blue points). Errorbars are plotted and in most cases are smaller than the symbol size. The errors reported (by GIPSY/ROTCUR) include only errors on the fits, while the systematic error considered ($3\degree$) is shown as a Green errorbar.}
\label{inclination}	 
\end{centering}
\end{figure*}

\begin{figure*}
\begin{centering}
\includegraphics[angle=0,width=10truecm]{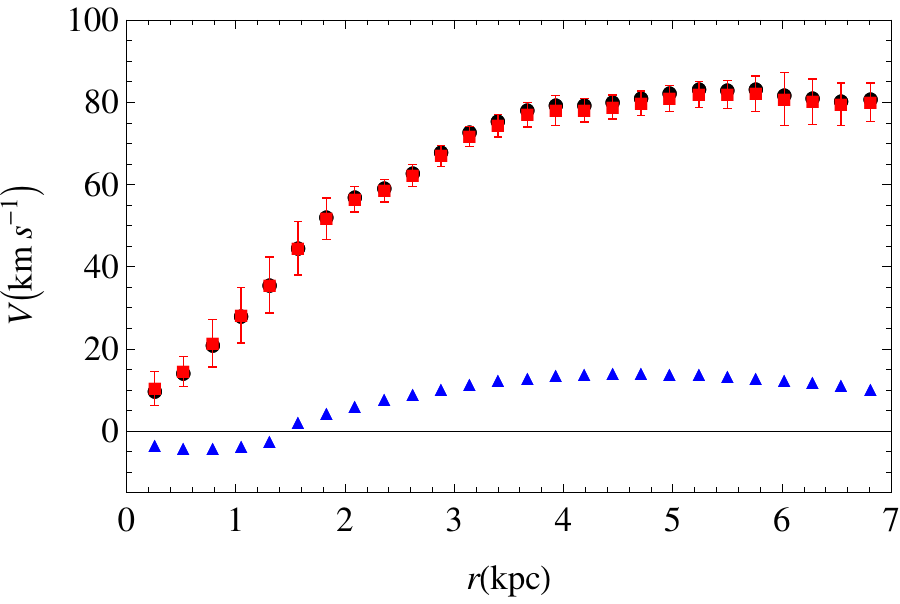}
\caption{Rotation velocity corrected for asymmetric drift (filled black
  points), raw velocity data with error bars (filled red boxes) and
  the asymmetric drift correction (filled blue triangles).}
\label{asymm}	
\end{centering}
\end{figure*}

The channel maps of the Orion dwarf provide evidence of well-ordered
rotation throughout the \hi\ disk (see {Cannon et
  al. 2010}\nocite{Cannon}).  The intensity-weighted-mean velocity
field (Figure (\ref{figcap3}b)) exhibits symmetric structure in the
outer disk.  Twisted iso-velocity contours at inner radii coincide
with the \hi\ holes near the centre of the disk.  The disk is
therefore dominated by circular motion. The RC of the galaxy was
derived by fitting a tilted ring model to the intensity-weighted-mean
velocity field using the {\sc gipsy} task ROTCUR.  The routine carries
out a least-squares fit to $V_{los}$,  the line of sight-velocity.
%
%\begin{equation}
%V_{los}(x,y)=V_{sys}+V_{rot}\sin i\cos\theta,
%\end{equation}
%
%where $V_{los}$ is the line of sight-velocity at a position on the sky
%with rectangular coordinates (x,y).  The angle from the major axis in
%the galaxy plane, $\theta$, is related to the sky-plane position
%angle, PA, by
%
%\begin{eqnarray}
%\cos\theta&=&\frac{-(x-X_c)\sin PA+(y-Y_c)\cos PA}{r},\\
%\sin\theta&=&\frac{-(x-X_c)\cos PA + (y-Y_c)\sin PA}{r\cos i},
%\end{eqnarray}
%
%where $(X_c, Y_c)$ are the coordinates of the dynamical centre and $r$
%is the galactocentric radius of the ring in the galaxy plane.  
To derive the best-fitting model, an iterative approach was adopted in
which the various combinations of the parameters were fitted.  The
final RC was extract by fixing all other parameters.  The receding and
approaching sides of the galaxy were fitted separately.  The best
fitting parameters are $i=(47\pm3)\degree$, $P.A. = (20\pm2)\degree$,
$V_{\rm sys} = 368.5\pm1.0$~km/s, and
($\alpha_{2000}$,$\delta_{2000}$) = (05:45:01.66, 05:03:55.2) for the
dynamical centre. We have realized that the inclination is not dependent on the radius, and the fit is shown in Figure (\ref{inclination}). Its weighted value is $(46.8 \pm 0.14)\degree$. Notice that because the errors reported by GIPSY/ROTCUR include only errors on the fits and systematics are not included, the $3\degree$ error estimate comes from attempting the ROTCUR fits in various orders (e.g., holding each variable fixed in turn).\\
The resulting RC is shown in Figure
(\ref{asymm}). Notice that in this object the disk inclination is
determined kinematically and therefore it is quite accurate. No result of this paper changes by adopting different values of  $i$, inside the  quoted errorbar.

The second-order moment map for the galaxy is shown in Figure
(\ref{figcap3}c). Throughout most of the disk, the velocity dispersion
is roughly constant at $\sigma$ $\simeq$ 7\,$\pm$\,2 km/s, with a more
complex behaviour near the galaxy centre and at the outermost radii.
This velocity dispersion estimate allows us to derive the asymmetric
drift correction to the RC yielded by the tilted ring model.  The
observed rotation velocity, $V_{rot}$, is related to the circular
velocity $V_c$ via
\beq
V^2_{c}(r)= V^2_{rot}(r) -  \sigma^2(r)\  \left[\frac{dlog \ \mu_{\rm HI}(r)}{dlog \ r} + \frac{dlog \  \sigma^2(r)}{dlog\  r}\right].  \label{circ}
\eneq

From an examination of Figure (\ref{asymm}) it is clear that the
$V_{rot}$ and $V_c$ profiles differ by less than 1$\%$. Throughout
this paper, we use the latter for the purposes of mass modelling. We
notice that in very small dwarfs this correction is not negligible ($V_{rot}\sim\sigma$) and it
introduces an uncertainty in the analysis, e.g. \citet{begum}.
  
In summary, the Orion dwarf RC has a spatial resolution of 0.26 kpc
(i.e. 0.2 $R_D$), and extends out to 5.1 R$_D$.  The uncertainties on
the RC are few km/s and the error on the RC slope $dlogV/dlogR<0.1$.
    
Is the circular velocity given by eq. (\ref{circ}) a proper estimate
of the gravitational field?  To further investigate the presence of
non-circular motions within the \hi disk that jeopardize the
kinematics, we carried out a harmonic decomposition of the
 intensity-weighted velocity field to search for any significant non-circular
components. This test is necessary in that the undetected presence of
non-circular motions can lead to incorrect parametrization of the
total mass distribution. \\  
Following \citet{shoen}, the line-of-sight velocities from the
\hi\ velocity field are decomposed into harmonic components up to
order $N=3$ according to
\begin{equation}
V_{los}=V_{sys}+\sum_{m=1}^{3} c_m  \cos m\theta+s_m  \sin m\theta, 
\end{equation}
\noindent where $V_{sys}$ is the systemic velocity, $c_m$ and $s_m$
are the magnitudes of the harmonic components, $m$ the harmonic
number, and $\theta$ the azimuthal angle in the plane of the galaxy.
The {\sc gipsy} task RESWRI was used to carry out the decomposition by
fitting a purely circular model to the velocity field, subtracting it
from the data, and then determining from the residual the magnitudes
of the non-circular components.  The tilted ring model fitted by
RESWRI had its kinematic centre fixed to that of the purely circular
tilted ring model used to derive the RC above.  The position angles
and inclinations were fixed to constant values of 20 and $47\degree$,
respectively.

\begin{figure*}
\begin{centering}
\includegraphics[angle=0,width=2\columnwidth]{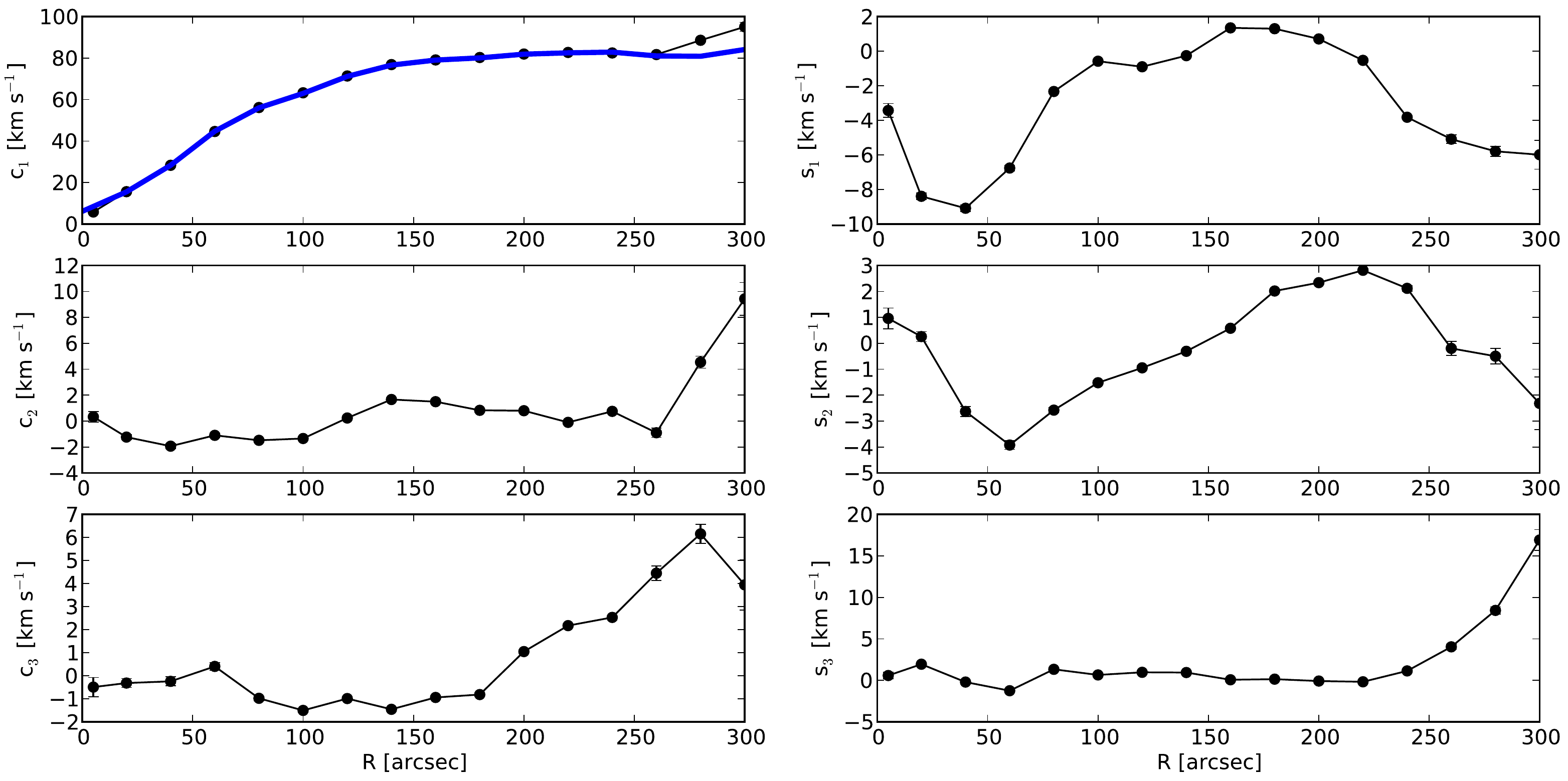}
\caption{Non-circular velocity components as derived by carrying out
  a harmonic decomposition of the intensity-weighted-mean velocity
  field.  The $c_1$ profile corresponds to the circular RC.  The RC
  derived by fitting tilted ring model to the intensity-weighted-mean
  velocity field is shown as a solid blue curve. }
\label{reswri4}	
\end{centering}
\end{figure*}

The parameters of the best-fitting model are shown in Figure
(\ref{reswri4}). Adjacent points are separated by a beam width in order  to ensure that they are largely independent of one another.  We argue that because the standard tilted ring model has fewer free parameters than the model incorporating the higher order Fourier components, it is not as essential to space the points on the rotation curve by a full beam width. Then in this model only $16$ points are considered instead of the $26$ points used in fitting the RC. \\
At inner radii the inferred non-circular motions are
not negligible, but this is almost certainly due to the fact that the
\hi\ distribution over this portion of the disk is irregular, being
 dominated by the large central \hi\ under-densities.  The
harmonic components of the outer disk are, instead, reliable and
demonstrate the gas flow to be dominated by circular kinematics.  The
circular velocity so obtained well matches that found by means of the
tilted ring model presented above.  The amplitudes of $c_2$ and
$s_2$ are too small to hide a cusp inside an apparently solid body RC
(as suggested by \citet{haya}).  These results provide further
decisive support for the use of $V_{rot}$ of the Orion dwarf as a tracer of its
mass distribution.

\section{Mass modeling}

We model the Orion dwarf as consisting of two ``luminous'' components,
namely the stellar and the gaseous disks, embedded in a dark halo. The
stellar component is modelled as an exponential thin disk \citep{freeman} with a scale
length of 1.33~kpc.  Any bulge component is assumed to be negligible
in terms of mass.  The dynamical contribution of the gas to the
observed RC is derived from the \hi\ total intensity map.  A scaling
factor of 1.33 is incorporated to account for the presence of Helium
and other elements.  For the dark halo we consider two different
parametrizations of the mass distribution: an NFW profile \citep{nfw}
and the cored profile of the Halo Universal Rotation Curve (URCH) \citep{salII}. It is well known that
the NFW profile is one outcome in numerical simulations of cold dark
matter structure formation, whereas the cored profile (an empirical
result), by design, fits the broad range of RC shapes of spiral
galaxies.
  
\subsection{Mass Models: stellar disk + dark matter halo}

The RC is modelled as the quadrature sum of the RCs of the individual
mass components:

\begin{equation}
V^2_{mod}=V^2_D+V^2_{DM}+V^2_{\rm gas}.
\end{equation}
For the cored halo parametrization we adopt the URCH profile:
\bea
V^2_{URCH}(r)&=&6.4\frac{\rho_0r_0^3}{r}\left[
\ln\left(1+\frac{r}{r_0}\right)-\arctan\left(\frac{r}{r_0}\right)       \label{eq:1.0}
\right. \nonumber \\
&+& \left. \frac{1}{2}\ln\left(1+\frac{r^2}{r_0^2}\right)\right],                        
\enea
where the disk mass, the core radius $r_0$ and the central halo
density $\rho_0$ are free parameters.

\begin{figure*}
\begin{centering}
\includegraphics[width=8cm,height=5cm]{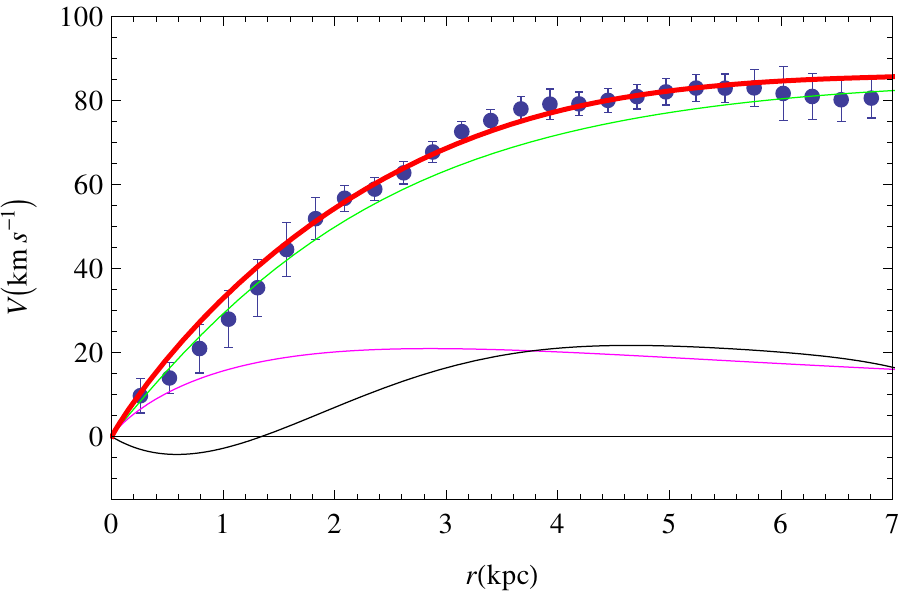}
\includegraphics[width=8cm,height=5cm]{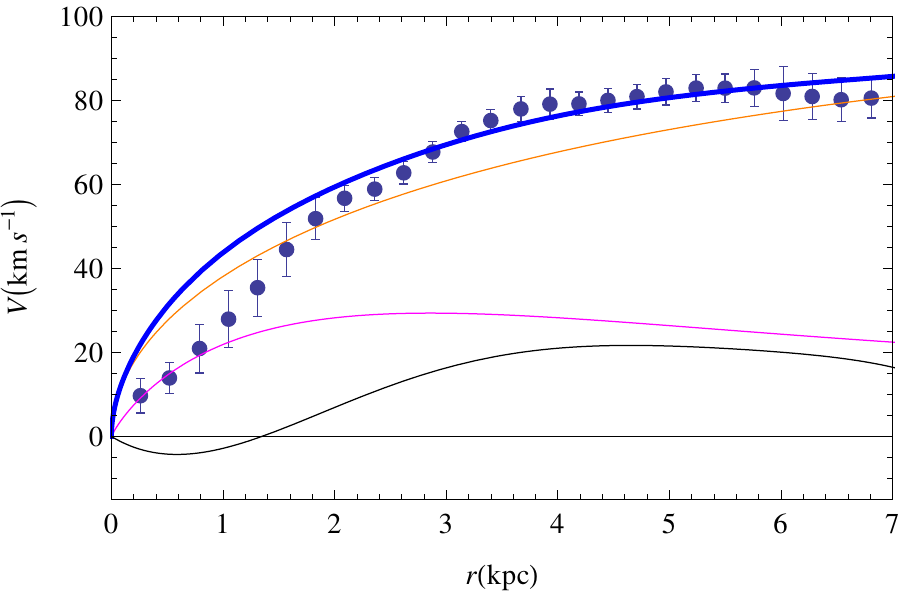}
\caption{Left: URC model of Orion dwarf. The circular velocity (filled
  circles with error bars) is modeled (thick Red line) by a Freeman
  disk (Magenta line), a URCH halo (Green line), and the \hi\ circular
  velocity (Black line).  Right: NFW model of Orion dwarf. The
  circular velocity data (filled circles with error bars) is modeled
  (thick Blue line) including a Freeman disk profile (Magenta line), a
  NFW halo profile (Orange line) and the \hi\ circular velocity (Black
  line).  In both cases the values of the free parameters are reported
  in the text.}
\label{URCNFW}
\end{centering}
\end{figure*}

It is evident that this model yields a total RC that fits the data
extremely well (see Figure (\ref{URCNFW}) left panel), with
best-fitting parameters of $r_0=(3.14\pm0.32)$ kpc, M$_D=(3.5\pm
1.8)\times10^8$ M$_\odot$\ and $\rho_0=(4.1\pm
0.5)\times10^{-24}$g/cm$^3$. More accurate statistics is not
necessary; the mass model predicts all the $V(r)$ data points within
their observational uncertainty.

Notice that the derived value of the disk mass
agrees with the photometric estimate discussed above.  The
corresponding virial mass and radius of the DM halo are M$_{vir}=(5.2
\pm 0.5) \times10^{10}$ M$_\odot$ (see eq. 10 in \citet{salII}) and
R$_{vir}=95.5^{+5}_{-4} $ kpc \citep{eke}, respectively.  We note that the
Orion dwarf has a mass 20 times smaller than that of the Milky Way,
with the DM halo dominating the gravitational potential at all
galactocentric radii. The baryonic fraction is $f_b = (M_D+M_{\rm
  gas})/M_{vir}=0.016$, while the gas fraction is $M_{\rm gas}/M_{vir}=
9\times10^{-3}$.

The RC for the NFW dark matter profile is 
\beq
V^2_{NFW}(r)=V^2_{vir}\frac{g(c)}{xg(cx)},
\eneq
where $x=r/R_{vir}$, $g(c)=\left[ln(1+c)-c/(1+c) \right]^{-1}$ and $c(M_{vir})=9.60\left(M_{vir}/10^{12}h^{-1}M_{\odot}\right)^{-0.075}$ is the concentration parameter  (see \citet{klypin}).  We fitted the RC of
the Orion dwarf by adjusting M$_{vir}$ and M$_{D}$.  The resulting
best-fit values are M$_{vir}=(2.5\pm0.5)\times 10^{11}$ M$_\odot$ and
M$_D=(6.9\pm1.7)\times10^8$ M$_\odot$, but since $\chi_{\mbox{\small red}}^2\simeq 3.3$, i.e. the fit is unsuccessful, the best-fit values of the free parameters and those of their fitting uncertainties do not have a clear  physical meaning.  
We plot the results in the right panel of Figure (\ref{URCNFW}).  The NFW model, at
galactocentric radii r$<$2 kpc, overestimates the observed circular velocity (see Figure (\ref{URCNFW}) right panel).

\section {Orion kinematics in the  MOND context}

An alternative to Newtonian gravity was proposed by \citet{milgroma}
to explain the phenomenon of mass discrepancy in galaxies.  It was
suggested that the true acceleration $a$ of a test particle, at low
accelerations, is different from the standard Newtonian acceleration,
$a_N$:

\beq
a=\frac{a_N}{\mu(a/a_0)},
\eneq
where $\mu(a/a_0)$ is an interpolation function and
$a_0=1.35\times10^{-8}$cm s$^{-2}$ is the critical acceleration at
which the transition occurs (see \citet{famaey}).  For this we adopt the following form of
the interpolation function (see \citet{famaey2}):
\beq
\mu(a/a_0)=\frac{a/a_0}{1+a/a_0}. 
\eneq 
 
In this framework the circular velocity profile can be expressed as a
function of $a_0$ and of the standard Newtonian contribution of the
baryons to the RC, V$_{bar}$=(V$^2_{D}$+V$^2_{\rm gas}$)$^{1/2}$, obtaining for it
\beq
V^2_{MOND}=V^2_{bar}(r)\left(\frac{\sqrt{1+\frac{4a_0r}{V^2_{bar}(r)}}+1}{2}\right), \label{mond}
\eneq
(see \citet{rich}). 
Eq. (\ref{mond}) shows that in the MOND framework the resulting RC is
similar to the no-DM standard Newtonian one, with an additional term
that works to mimic and substitute for the DM component \citep{sanders02}. No result of this paper changes by adopting the ``standard''  MOND  interpolation function (see \citet{zhao}). 

\begin{figure*}
\begin{minipage}[]{8.5cm}
\centering
\vspace{0.3cm}
\includegraphics[width=8cm]{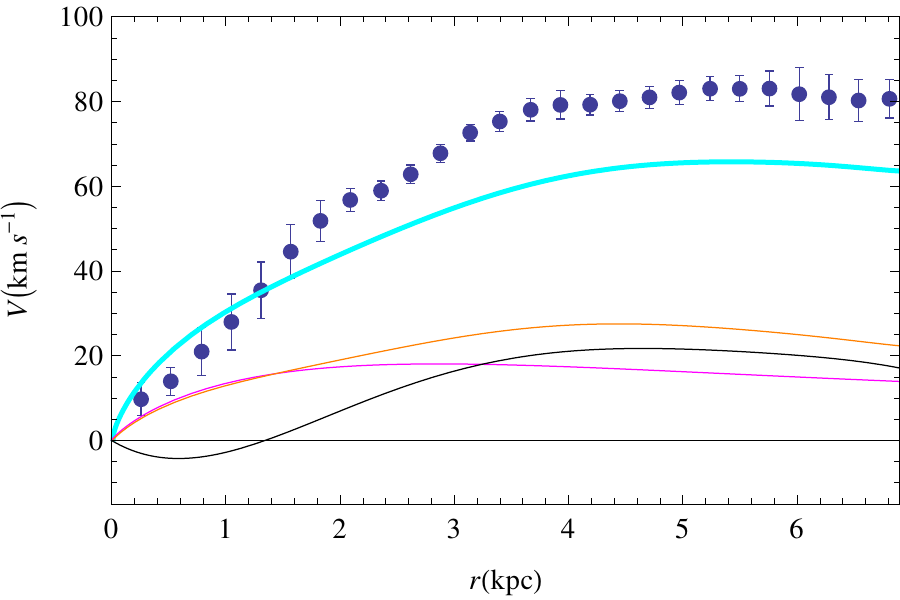}
\caption{MOND model of Orion dwarf spiral. The circular velocity data
  (filled circles with error bars) is modeled with the MOND profile
  (thick Cyan line). Also shown are the (newtonian) Freeman disk
  contribution (Magenta line), the \hi\ contribution (Black line) and
  the total baryonic contribution (Orange line).}
\label{MOND1}
\end{minipage}
\ \hspace{3mm} \hspace{3mm} \
\begin{minipage}[]{8cm}
\centering
\includegraphics[width=8cm]{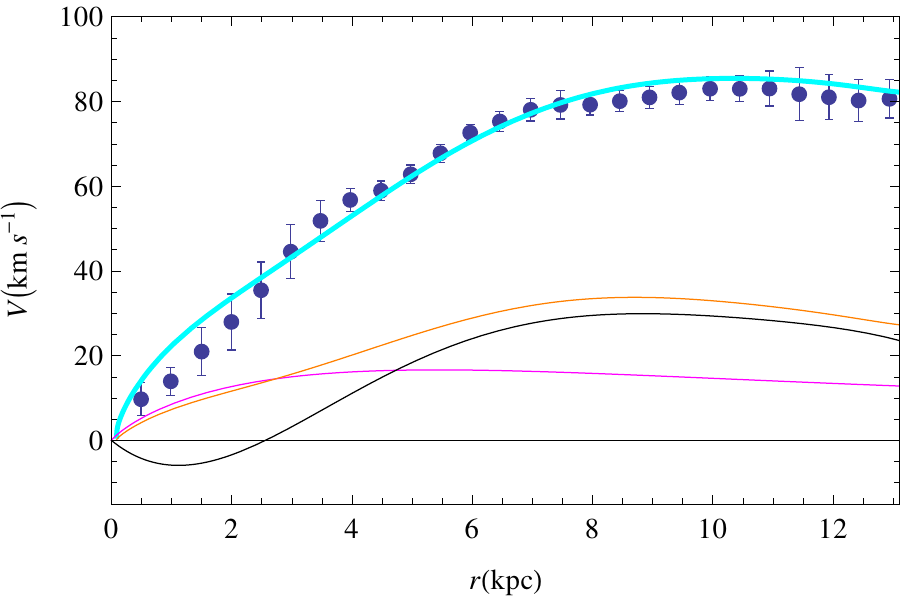}
\caption{MOND model of Orion dwarf spiral for a distance from the
  galaxy of $\sim$10 Mpc, a factor 1.9 farther than the nominal
  distance. The legend to the lines is the same of
  Figure \ref{MOND1}. The resulting disk mass is M$_D=4.2\times10^8$
  M$_\odot$.}
 \label{MOND2}
\end{minipage}
\end{figure*}

The best-fitting MOND mass model is shown in Figure (\ref{MOND1}). The
model total RC (cyan line) completely fails to match the
observations. We fix the stellar mass M$_D$ at 
M$_D=2.6\times10^8$\msun. If we let the disk mass becomes higher, covering the mass range estimated in \citet{Cannon}, the fit is not even able to reproduce the RC at inner radii. Note that in the MOND formalism, the distance of
the galaxy and the amount of gaseous mass are both crucial in deriving
the model RC. To quantify the discrepancy of these observations with
the MOND formalism, note that only if the Orion dwarf were 1.9 times
more distant than the current estimate we would obtain a satisfactory
fit to the RC (see Figure (\ref{MOND2})). 

%Similarly, a doubling of the
%distance would yield a satisfactory RC fit, if one allows the
%\hi\ mass to have a larger value (e.g., by accounting for
%\hi\ self-opacity; see {Braun et al. 2009}\nocite{braun}), in other words $\left|\frac{dM_{\rm HI}}{M_{\rm HI}}\right|+2\left|\frac{dD}{D}\right|\simeq3$.

\section{Conclusions}

The Orion dwarf galaxy is representative of a population of dwarfs
with a steep inner RC that gently flattens at the edge of the
gas disk.  The observed kinematics imply the presence of large
amounts of DM also in the central regions.  We have used new
\hi\ observations of the Orion dwarf to analize its kinematics and
derive the mass model. The derived RC is very steep and it is
dominated by DM at nearly all galactocentric radii. Baryons are unable
to account for the observed kinematics and are only a minor mass
component at all galactic radii.  

We have used various mass modeling approaches in this work.  Using the
NFW halo, we find that this model fails to match the observed
kinematics (as occurs in other similar dwarfs). We show that
non-circular motions cannot resolve this discrepancy.  Then we modelled the galaxy by assuming the URCH parametrization of the
DM halo.  We found that this cored distribution fits very well the
observed kinematics. Orion is a typical dwarf showing a cored profile
of the DM density and the well-known inability of DM halo cuspy
profiles to reproduce the observed kinematics.  Finally, we find that the MOND model is discrepant with the
data if we adopt the literature galaxy distance and gas mass. The
kinematic data can be reproduced in the MOND formalism if we allow for
significant adjustments of the distance and/or value of the gas mass.
Let us point out that the present interferometric observations may miss some of the objects' \hi flux,
although this may be limited in that the cubes do not have significant negative bowls. Obviously, for  bigger values of the \hi mass, the distance at which the baryon components would well fit the data will also somewhat  decrease.\\
It is worth stressing that there is a  galaxy distance (albeit presently not-favoured) for which MOND would  strike an extraordinary success in reproducing the observed kinematics of the Orion dwarf.  

The Orion dwarf has a favorable inclination, very regular gas kinematics, a small asymmetric drift correction, a well-understood baryonic matter distribution, and a large discrepancy between luminous and dynamical mass.  All of these characteristics make this system a decisive benchmark for the MOND formalism and a promising target for further detailed studies.  Of particular value would be a direct measurement of the distance (for example, infrared observations with the Hubble Space Telescope would allow a direct distance measurement via the magnitude of the tip of the red giant branch).

\section*{Acknowledgments}

The authors would like to  thank the referee, Gianfranco Gentile, for his very fruitful comments that have increased the level of presentation of the paper.

\label{lastpage}
\end{document}